\documentclass[letterpaper]{article} 
\usepackage{aaai2026}  
\usepackage{times}  
\usepackage{helvet}  
\usepackage{courier}  
\usepackage[hyphens]{url}  
\usepackage{graphicx} 
\usepackage{natbib}  
\usepackage{caption} 
\usepackage{color}
\usepackage{amsmath}
\usepackage{amsthm}
\usepackage{multirow}
\usepackage{booktabs}
\usepackage[ruled,linesnumbered]{algorithm2e}
\usepackage{float}
\usepackage{amsfonts}

\newtheorem{theorem}{Theorem}  

\title{TransFR: Transferable Federated Recommendation with Adapter Tuning on Pre-trained Language Models}
\author{
    Honglei Zhang\textsuperscript{\rm 1,2}, Zhiwei Li\textsuperscript{\rm 3}, Haoxuan Li\textsuperscript{\rm 4}, Xin Zhou\textsuperscript{\rm 5}, Jie Zhang\textsuperscript{\rm 5}, Yidong Li\textsuperscript{\rm 1,2}\thanks{Corresponding author.}
}
\affiliations{
    \textsuperscript{\rm 1}Key Laboratory of Big Data \& Artificial Intelligence in Transportation, Ministry of Education, China\\
    \textsuperscript{\rm 2}Beijing Jiaotong University, China\\
    \textsuperscript{\rm 3}University of Technology Sydney, Australia \\
    \textsuperscript{\rm 4}Peking University, China\\
    \textsuperscript{\rm 5} Nanyang Technological University, Singapore\\
    \{honglei.zhang, ydli\}@bjtu.edu.cn

}

\usepackage{bibentry}

\begin{document}

\maketitle

\begin{abstract}
Federated recommendations (FRs), facilitating multiple local clients to collectively learn a global model without disclosing user private data, have emerged as a prevalent on-device service. In conventional FRs, a dominant paradigm is to utilize discrete identities to represent clients and items, which are then mapped to domain-specific embeddings to participate in model training. Despite considerable performance, we reveal three inherent limitations that can not be ignored in federated settings, i.e., non-transferability across domains, ineffectiveness in cold-start settings, and potential privacy violations during federated training. To this end, we propose a transferable federated recommendation model, \textit{TransFR}, which delicately incorporates the general capabilities empowered by pre-trained models and the personalized abilities by fine-tuning local private data. Specifically, it first learns domain-agnostic representations of items by exploiting pre-trained models with public textual corpora. To tailor for FR tasks, we further introduce efficient federated adapter-tuning and post-adaptation personalization, which facilitate personalized adapters for each client by fitting local private data. We theoretically prove the advantages of incorporating adapter tuning in FRs regarding both effectiveness and privacy. Through extensive experiments, we show that our TransFR surpasses state-of-the-art FRs on transferability.
\end{abstract}


\section{Introduction}

\begin{figure}[t]
\begin{center}
\includegraphics[scale=0.43]{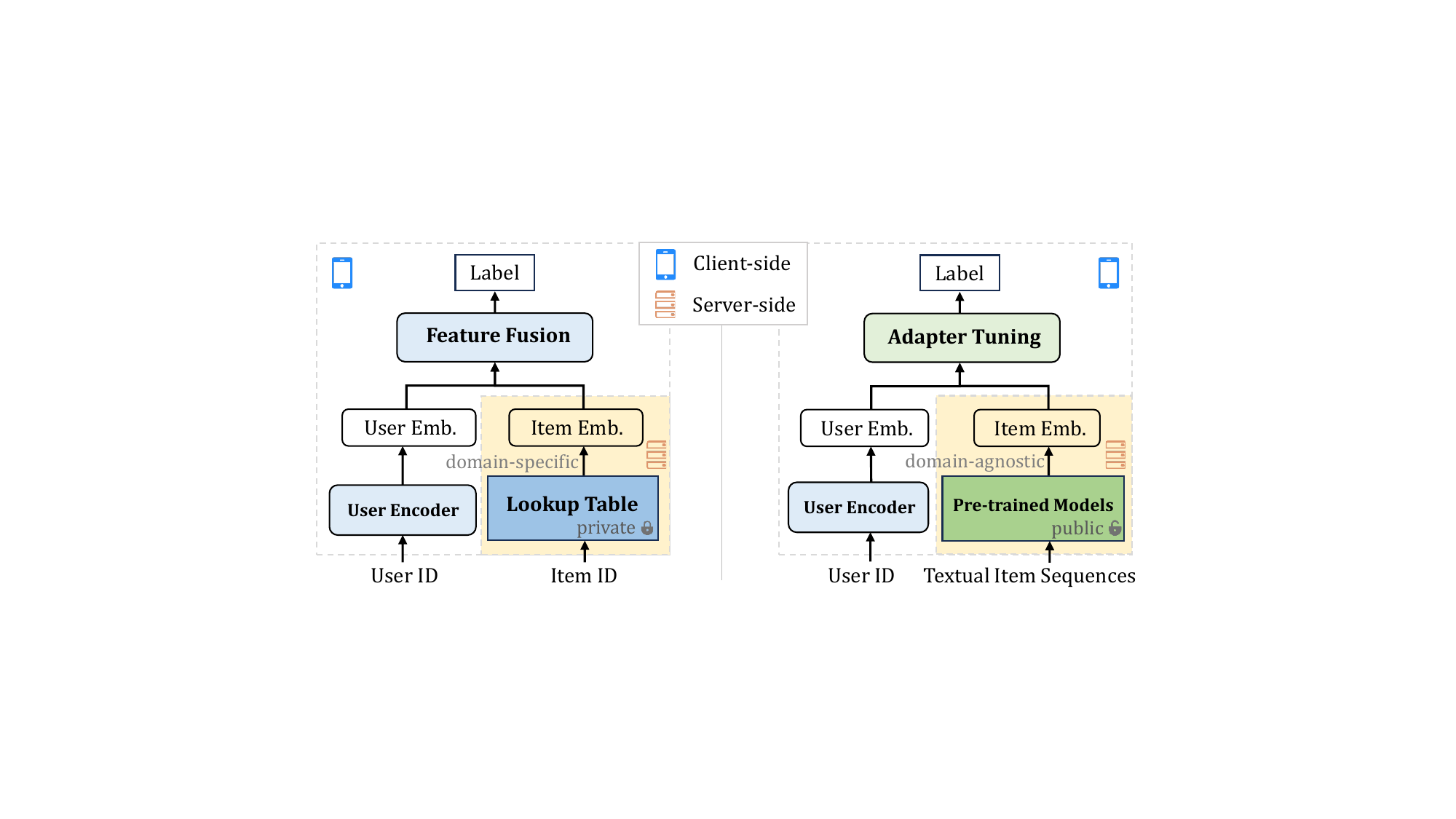}
\end{center}
\caption{Illustration of the differences between standard FR (left) and our TransFR model (right). Existing work operates as a lookup table with domain-specific item embeddings, while TransFR exploits pre-trained language models to produce universal domain-agnostic embeddings.
}
\label{fig:intuition}
\end{figure}

Federated recommendations (FRs), owing to the benefits of efficient and secure distributed learning~\cite{feng2018privacy}, have recently attracted a lot of interest both in academia and industry~\cite{zhang2025personalized,du2024enhancing,wu2022leveraging}. Empirically, it can be ingeniously generalized into a unified FR framework~\cite{zhang2024beyond,zhang2024privfr,chen2025breaking}, which initially involves a user encoder and a fusion module in local client, and an item encoder in central server, as illustrated in Fig.~\ref{fig:intuition}. Many canonical works have expanded this research line in three directions. For instance, HPFL explores learning a powerful user encoder~\cite{hpfl_2021}, RULE focuses on modeling efficient item embeddings~\cite{rule_2021}, and Zhang \textit{et al.} aims to design adaptive fusion modules~\cite{pfedrec_2023}. Subsequent research has built upon this foundation by adding textual data into FR fields to mitigate the data sparsity issue~\cite{li2025personalized,hu2025modality}. For instance, IFedRec incorporates the textual descriptions of items based on user interactions to improve model performance~\cite{ifedrec2024}. It is worth noting that all these federated models mainly utilize discrete identities (IDs) to represent items, which have been an effective solution within single domains.



Although ID-based FR models have been well established, they routinely suffer from the following multilevel challenges in cross-domain tasks. First, it lacks \textit{transferability} across platforms, as item IDs are generally not shared in practical scenarios. To be specific, the same IDs across different domains may have different meanings, hence the embeddings trained with these IDs are domain-specific, which seriously limits the development of transferable FR models~\cite{fedfast_2020}. Second, it fails to handle the \textit{cold-start issue} due to the ID-based item embeddings. Specifically, a well-trained embedding highly relies on sufficient ID interactions, which cannot provide satisfactory results when users and items have few interactions~\cite{ifedrec2024}. Third, it poses a \textit{privacy breach issue} in FR settings since the item embeddings are trained directly with user private data~\cite{zhang2023feddcsr}. In summary, none of these FR models take into account all the issues, which are the primary challenges for real-world FRs.



Given the challenges of existing work, we believe it is essential to develop a universal FR model that can effectively transfer to a new domain while ensuring unseen items can be handled well in cold-start settings. To achieve this, we can simply resort to the notion of retraining techniques to train the model from scratch with new domain-specific data~\cite{retrain_2020}. Besides, we can augment side information or perturbations during the retraining process to alleviate the cold-start and privacy issues, respectively~\cite{ifedrec2024,frs_dp_2020}. However, such a straight solution brings a significant efficiency bottleneck since retraining the model involves multiple rounds of communication in federated learning~\cite{guo2025federated,feng2025sadba}, especially in resource-limited cross-device tasks. Moreover, retraining implies directly accessing to local raw data in the new domain, undoubtedly increasing the risk of user privacy leakage. Hence, it is imperative to design an efficient transferable FR framework, which can represent items in a domain-agnostic manner, following a privacy-preserving paradigm.


Inspired by recent findings suggesting that ID features of items can be replaced by textual features~\cite{idfree}, we present \emph{TransFR}, a principled transferable FR framework built on foundation models and federated adapter tuning techniques. Specifically, it represents items in a domain-agnostic way by purely utilizing fine-grained textual embeddings (as shown in the right panel of Fig.~\ref{fig:intuition}) instead of discrete item IDs. During this phase, language models like BERT are trained on publicly available text corpora on resource-rich servers. It can be understood that before the enactment of GDPR\footnote{https://gdpr-info.eu/}, these models could be trained with centrally collected data to achieve superior performance. After that, clients conduct efficient federated adapter tuning process, adapting it to the recommender task without transmitting raw data. This phase can be seen as post-GDPR period, adhering to legal restrictions on data stored at local clients. Finally, we introduce post-adaptation personalization to further enhance the personalization of FR tasks. Overall, our main contributions are listed as follows:

\begin{itemize}
\item We propose a unified FR model, TransFR,  which can delicately exploits universal capabilities from foundation models and the personalized abilities from fine-tuning techniques. To the best of our knowledge, this is the first work focusing on textual transferability in FRs.

\item To tailor for FRs, we introduce a simple yet effective federated adapter tuning to endow generic language models with collaborative signals. We further prove its effectiveness and privacy of adapter tuning in FRs theoretically, ensuring a dual guarantee of efficiency and privacy.


\item To achieve personalized FRs, we introduce a post-adaptation personalization module, with locally fitting a prediction head on private data distributions, effectively mitigating data heterogeneity across clients. 


\item Extensive experiments on several datasets show the advantages of our model on effectiveness and transferability over state-of-the-art counterparts.

\end{itemize}


\section{Related Work}

In this section, we review two relevant areas to this work, \textit{i.e.}, federated recommendation~\cite{zhang2025personalized} and transferable recommendation~\cite{zang2022survey}. 


\subsection{Federated Recommendation}




Federated recommendation (FR) has emerged as a promising mobile service in recent years, due to its ability to preserve user privacy~\cite{li2025learn,yi2025pfedes,chen2025beyond}. FCF~\cite{fcf_2019} and FedRec~\cite{fedrec_2020} are two pioneering works to utilize the discrete identities (IDs) to learn user/item embeddings. In general, there are three major components in the standard FR framework, i.e., user encoder, item encoder, and fusion module~\cite{pfedrec_2023}. As for the user encoder, Luo \textit{et al.} proposed a personalized user encoder by combining the global and local clustered models~\cite{perfedrec_2022}. As for the item encoders, researchers have been working to address the privacy~\cite{frs_dp_2020} and efficiency~\cite{lightfr_2022} issues associated with representing item embeddings with IDs. As for the fusion module, FedCIA proposed a similarity-based model fusion strategy to prevent information loss during the aggregation process~\cite{fedcia2025}. Recent work, IFedRec~\cite{ifedrec2024}, leveraged item attributes as side information to alleviate the cold-start issue in FR framework. Different from relying on coarse-grained IDs, our work represents items with fine-grained textual tokens to alleviate the transferability, cold-start and privacy issues of FRs.

\subsection{Transferable Recommendation}

Transferable recommendation aims to improve target-domain performance by leveraging auxiliary data from other domains, mitigating data sparsity issues~\cite{zang2022survey,bao2022minority,zhang2024uncovering}. Early approaches relied on explicit or implicit domain overlaps, using mapping functions for knowledge transfer~\cite{embeddingcr_2021}. For instance, EMCDR used a multi-layer perceptron for mapping to improve the robustness of target domain\cite{embeddingcr_2021}. Distinct from the above work, subsequent efforts focus on learning universal representations with language models across different domains~\cite{morec_2023}. For example, UniSRec utilized the associated descriptions of items to learn transferable embeddings across different scenarios, showing the transferability in a cross-platform setting~\cite{unirec_2022}. Subsequent works aim to enhance transferability in federated settings. For example, FedDCSR adopted disentangled ID-based embeddings for transfer, but may expose sensitive data during embedding transmission~\cite{zhang2023feddcsr}. FedCLR infered user interaction distributions in the source domain, reconstructs target-domain interactions via a generative model~\cite{wang2025federated}. Unlike prior methods that transfer item embeddings tied to private user data, our approach first learns transferable public parameters from open corpora, and then applies adapter tuning with private data, ensuring both privacy and transferability.


\section{The Proposed TransFR Framework}

In this section, we introduce the details of TransFR model.

\begin{figure}[!thp]
\begin{center}
\includegraphics[scale=0.37]{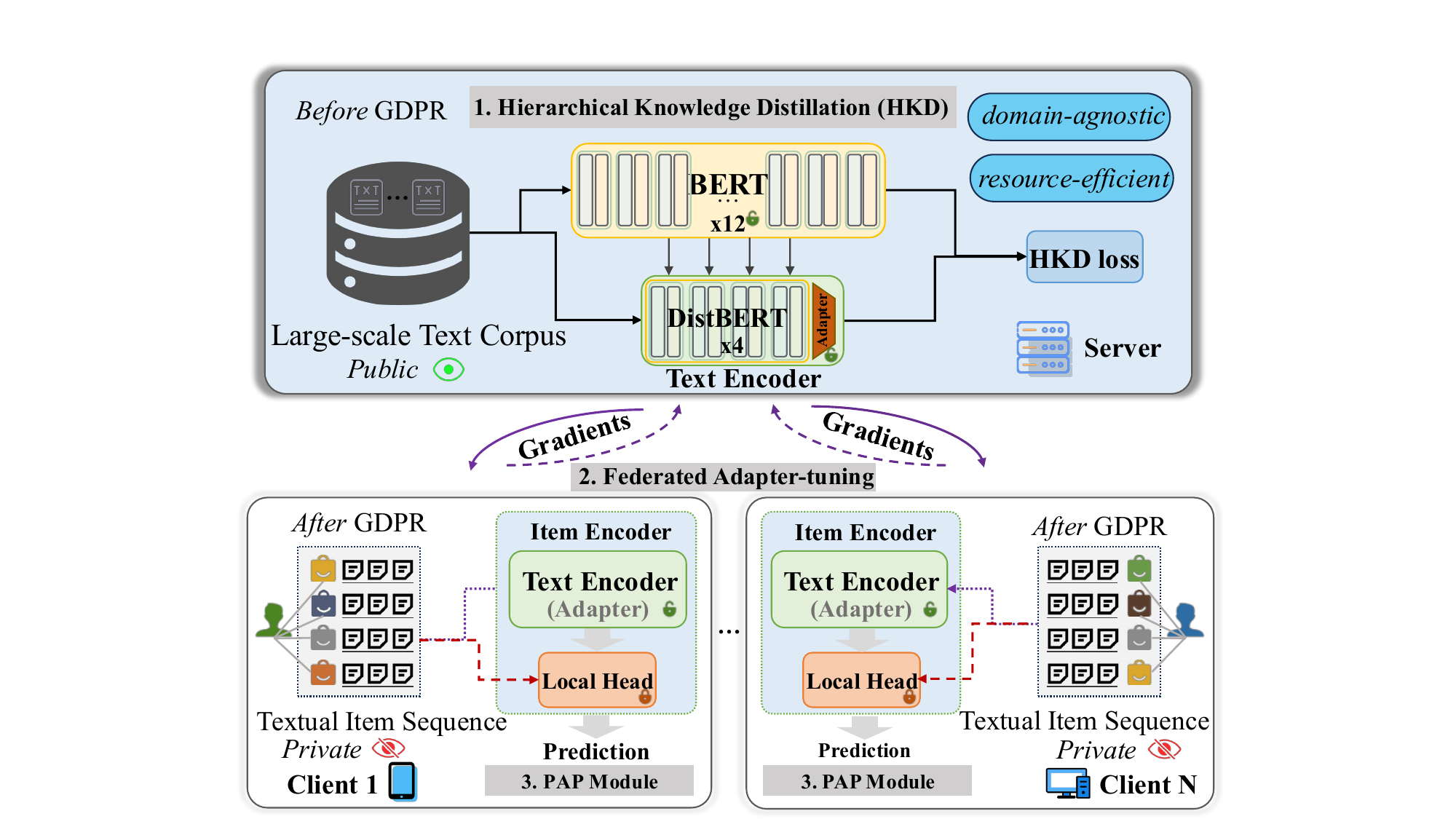}
\end{center}
\caption{The schematic diagram of TransFR, which includes hierarchical knowledge distillation, federated adapter-tuning and post-adaptation personalization (PAP).}
\label{fig:transfr}
\end{figure}

\subsection{Preliminary}


We consider a standard federated recommendation scenario with $n$ users $\mathcal{U}=\{u\}$ and $m$ items $\mathcal{I}=\{i\}$ on a central server. Each user $u$ holds local feedback data $\mathcal{D}_u$ consisting of interactions $(u, i, y_{ui})$, where $y_{ui}=1$ indicates user-item interaction. The goal is to predict $\hat{y}_{ui}$ for unseen items and recommend the top $K$ items to each user. In transferable recommendation, we define a source domain $S$ and multiple target domains $T=\{T_1, \dots, T_D\}$, each with their own user/item sets. In the source domain (and similarly in target domains), there exists a user set $\mathcal{U}^s$ ($\mathcal{U}^t$) and an item set $\mathcal{I}^s$ ($\mathcal{I}^t$), comprising $|\mathcal{U}^s|$ ($|\mathcal{U}^t|$) users and $|\mathcal{I}^s|$ ($|\mathcal{I}^t|$) items. Our work focuses on cross-domain item recommendation for shared clients, utilizing domain-agnostic textual embeddings as an exploration in federated settings. 



\subsection{The TransFR Model}


\subsubsection{Overview}

Fig.~\ref{fig:transfr} illustrates our framework. First, a central server pre-trains a large BERT model $\varphi_{\mathrm{BERT}}$ on public text data. To suit resource-limited devices, hierarchical knowledge distillation produces a smaller DistBERT $\varphi_{\mathrm{DistBERT}}$. DistBERT is then combined with an Adapter $\psi$ for efficient federated adapter-tuning, forming the Text Encoder $g = \psi \circ \varphi$. This encoder is deployed to user devices for local training. For personalization, a post-adaptation personalization module adds a local head $h_u$ per client, creating a personalized Item Encoder $\phi_i = h_u \circ g$. For the user representation $\mathbf{p}_u$ on local clients, we utilize local data to perform local training of the User Encoder $\phi_u$. Ultimately, we can utilize the fusion module $f(\cdot)$ to facilitate the prediction function for recommending item $i$ to user $u$, denoted as $\hat{y}_{ui} =f(\mathbf{p}_u, \mathbf{q}_i)$, where $\mathbf{q}_i$ denotes the produced embedding of item $i$. This process iterates until convergence.

\subsubsection{Server-side Execution Procedure.}

In TransFR, the server leverages its computational resources to pre-train and distill a universal language model from centralized text corpora. Given the availability of mature pre-trained models, we skip pre-training and directly use BERT~\cite{bert_2018} pretrained on BooksCorpus~\cite{bookscorpus2015} and Wikipedia. These datasets predate GDPR, permitting centralized use. We then apply hierarchical knowledge distillation to produce an efficient model for resource-limited devices.

\textbf{Hierarchical Knowledge Distillation.} To enable efficient deployment on end-user devices, we introduce a hierarchical knowledge distillation (HKD) method tailored for BERT~\cite{tinybert}. HKD transfers knowledge from a larger teacher BERT model with $l$ Transformer layers to a smaller student model with $s$ layers ($l > s$), both sharing the same architecture except for depth. The goal is to reduce model size and speed up inference while maintaining accuracy by minimizing a defined distillation loss.



\begin{equation}\label{eq:hkd}
\begin{gathered}
\mathcal{L}_{\mathrm{HKD}}=\sum_{x \in \mathcal{X}} \sum_{z=0}^{z} \mathcal{L}_{\mathrm{layer }}\left(\varphi^z_{\mathrm{DistBERT}}(x), \varphi^{p(z)}_{\mathrm{BERT}}(x)\right),
\end{gathered}
\end{equation}

\noindent where $x$ is the textual input and $\mathcal{X}$ the centralized training set collected before GDPR enforcement. The mapping function $l = p(z)$ maps multiple layers of the teacher model onto a single layer of the student model. Specifically, we set $p(z) = 3 \times z$, so DistBERT distills knowledge from every third teacher layer. The distillation loss targets individual BERT layers, primarily Transformer layers with multi-head attention. We design an attention-based loss that aligns the student’s attention matrices with the teacher’s, formally defined as follows:


\begin{equation}\label{eq:attention}
\begin{gathered}
\mathcal{L}_{\text {attn}}=\frac{1}{h} \sum_{i=1}^h \left(\boldsymbol{A}_i^{\operatorname{stu}}- \boldsymbol{A}_i^{\operatorname{tea}}\right)^2,
\end{gathered}
\end{equation}

\noindent where $h$ is the number of attention heads, and $\mathbf{A}_i \in \mathbb{R}^{L \times L}$ is the attention matrix of the $i$-th head, with $L$ as the input length. Inspired by studies showing BERT’s attention captures rich knowledge~\cite{bert_attention}, we use mean squared error loss to transfer knowledge from teacher to student. Besides attention, we also distill knowledge from the embedding layer, the objective is defined as:

\begin{equation}\label{eq:embedding}
\begin{gathered}
\mathcal{L}_{\text {embd}}=\left(\boldsymbol{E}^{\operatorname{stu}} \boldsymbol{W}- \boldsymbol{E}^{\operatorname{tea}}\right)^2,
\end{gathered}
\end{equation}

\noindent where the matrices $\mathbf{E}^{stu}\in\mathbb{R}^{L\times f}$ and $\mathbf{E}^{tea}\in\mathbb{R}^{L\times f^\prime}$ correspond to the embeddings of student and teacher models, respectively. $f$ and $f^\prime$ represent the embedding sizes for the student and teacher models, with $f$ typically being smaller than $f^\prime$ to achieve a more compact student model. The matrix $\mathbf{W}\in\mathbb{R}^{f\times f^\prime}$ is a learnable linear transformation aligning the dimensional space of the student network's embeddings with that of the teacher network's embeddings. Since in the following sections we primarily utilize the distilled smaller student model $\varphi_{\mathrm{DistBERT}}$, we will abusively use the symbol $\varphi$ to denote $\varphi_{\mathrm{DistBERT}}$ for clarity. 



\textbf{Efficient Federated Adapter-Tuning.} To adapt this pre-trained model for recommendation tasks, it is necessary to collaboratively fine-tune the model between the server and clients using local private data. To achieve this goal, we propose an efficient federated adapter-tuning (FAT) method, which introduces a small, learnable Adapter on top of the frozen model, effectively balancing privacy and efficiency. Specifically, we incorporate an Adapter $\psi_{\mathrm{Adapter}}$ into the DistBERT model $\varphi_{\mathrm{DistBERT}}$, resulting in a Text Encoder $g(\cdot)$ that explicitly combines both generality and personalization. The encoder $g(\cdot)$ can be formally represented as:




\begin{equation}\label{eq:text_encoder}
\begin{gathered}
g(\cdot)= \underbrace{\psi_{\mathrm{Adapter}}}_{\mathrm{trainable}} \circ \underbrace{\varphi_{\mathrm{DistBERT}}}_{\mathrm{frozen}},
\end{gathered}
\end{equation}

\noindent where the Adapter consists of multi-layer perceptrons. After the server aggregates the gradients $\mathbf{g}_{\theta}$ of Adapter $\psi$ in a weighted sum fashion uploaded from the sampled clients $\mathcal{U}_p$, the global aggregation and update of the Adapter $\psi$ as:


\begin{equation}\label{eq:aggregation}
\begin{aligned}
\theta_{\mathrm{adapter}} &=\theta_{\mathrm{adapter}} - \eta_s \cdot
\left(\frac{1}{|\mathcal{U}_p|}\sum\limits_{u\in\mathcal{U}_p} \mathbf{g}^u_{\theta}\right),
\end{aligned}
\end{equation}

\noindent where $\eta_s$ is the learning rate adopted in the server, and $\mathbf{g}_{\theta}^u$ represents the gradients of the Adapter uploaded from client $u$. This aggregation rule follows FedAvg~\cite{fl_first_2017}. We present more theoretical analysis on model effectiveness and privacy guarantee about our simple yet effective federated adapter tuning.The local optimization process will be detailed in the following section. 


\subsubsection{Client-side Execution Procedure.} In TransFR, clients handle local optimization for federated adapter-tuning by updating user embeddings and uploading adapter gradients for global aggregation. They also perform post-adaptation personalization with local data to enhance personalization.


\textbf{Efficient Federated Adapter-Tuning.} The clients mainly handle the local optimization steps during the efficient federated adapter-tuning. Given the user embedding $\mathbf{p}_u$ and item embedding $\mathbf{q}_i$ of item $i$ stored in client $u$, we adopt inner product to predict the preference score of user $u$ to item $i$ in the local device. Here, we utilize the well-designed binary cross-entropy (BCE) loss, which is as follows:


\begin{equation}\label{eq:bce}
\begin{aligned}
\mathcal{L}_{\mathrm{local}}=-\sum_{(u, i) \in \mathcal{D}_u} y_{u i} \log \hat{y}_{u i}+\left(1-y_{u i}\right) \log \left(1-\hat{y}_{u i}\right),
\end{aligned}
\end{equation}

\noindent where $\mathcal{D}_u=\mathcal{D}^+_u\cup \mathcal{D}^-_u$ and $\mathcal{D}^+_u$ denotes the set of observed interactions, \textit{i.e.}, $y_{ui}=1$, and $\mathcal{D}^-_u$ denotes the set of negative instances, \textit{i.e.}, $y_{ui}=0$, which can be uniformly sampled from unobserved interactions. For every observed interaction, there are $N$ negative instances sampled for efficient training. By optimizing the BCE loss in the local client, we can update $\mathbf{p}_u$ by stochastic gradient descent as follows:

\begin{equation}\label{eq:pu}
\begin{aligned}
\mathbf{p}_u&=\mathbf{p}_u - \eta_c \cdot \frac{\partial \mathcal{L}_{\mathrm{local}}}{\partial \mathbf{p}_u},
\end{aligned}
\end{equation}

\noindent where $\eta_c$ is learning rate in clients and the specific gradient form of $\frac{\partial \mathcal{L}}{\partial \mathbf{p}_u}$ is determined based on the adopted user encoder $\phi_u$. Besides, we calculate the gradients $\mathbf{g}_{\theta}=\{\Delta_{\theta_{\mathrm{bert}}},\Delta_{\theta_{\mathrm{adapter}}}\}$ of the local client for the global parameters $\theta_{\mathrm{enc}}=\{\theta_{\mathrm{bert}}, \theta_{\mathrm{adapter}}\}$ in Text Encoder $g(\cdot)=\psi \circ \varphi$, with the loss function $\mathcal{L}_{\mathrm{local}}$, which is as follows:

\begin{equation}\label{eq:local_gradient}
\begin{aligned}
\Delta_{\theta_{\mathrm{bert}}}=\frac{\partial \mathcal{L}_{\mathrm{local}}}{\partial \theta_{\mathrm{bert}}}, \quad \Delta_{\theta_{\mathrm{adapter}}}=\frac{\partial \mathcal{L}_{\mathrm{local}}}{\partial \theta_{\mathrm{adapter}}},
\end{aligned}
\end{equation}

\noindent where the parameters of the pre-trained BERT $\theta_{\mathrm{bert}}$ are frozen, hence $\Delta_{\theta_{\mathrm{bert}}}=0$ and $\mathbf{g}_{\theta}=\Delta_{\theta_{\mathrm{adapter}}}$. After the selected participating clients $\mathcal{U}_p$ completes the gradient calculation of global parameters, then only upload $\mathbf{g}_{\theta}$ to the server for global aggregation procedure.

\textbf{Post-Adaptation Personalization.} Inspired by recent pioneering work, which indicates that the success of centralized machine learning is often attributable to the shared global feature representation, while the heterogeneity across clients or tasks predominantly manifests in the labels~\cite{lecun2015deep}. To mitigate this heterogeneity, we introduce post-adaptation personalization (PAP) mechanism with a local prediction head $h_u(\cdot)$ on top of the Text Encoder $g(\cdot)$ for each client, which involves local training on the fine-tuned Item Encoder $\phi_i=h_u \circ g$ using private local data. This mechanism is designed to further fit the local data distributions, thereby facilitating personalized FRs. Specifically, we update the local prediction head using the local loss function $\mathcal{L}_{\mathrm{local}}$ in Eq.(\ref{eq:bce}), detailed below:

\begin{equation}\label{eq:local_training}
\begin{aligned}
\theta_{\mathrm{head}} &=\theta_{\mathrm{head}} - \eta_c \cdot
\frac{\partial \mathcal{L}_{\mathrm{local}}}{\partial\theta_{\mathrm{head}}},
\end{aligned}
\end{equation}

\noindent where $\theta_{\mathrm{head}}$ represents the trainable parameters of the local prediction head $h_u(\cdot)$, which is composed of multi-layer perceptrons. To reduce the risk of overfitting, the local head can adopt the same model structure as the Adapter, thereby reducing the learnable parameters in the Item Encoder. Intuitively, adding a personalized prediction head to each common pre-trained model can expand the optimization space of the training parameters, thereby yielding improved personalized prediction performance.

From a theoretical standpoint, this PAP strategy can be explained using domain adaptation theory~\cite{ben2010theory}. Let $P_S$ denote the global source distribution (on which the shared encoder $g$ is trained), and $P_T$ denote the client-specific target distribution. Then, the generalization error on the client domain can be bounded by:

\begin{equation}
\epsilon_T(h_u \circ g) \leq \epsilon_S(g) + \frac{1}{2}\mathcal{D}_{\mathcal{H}}(P_S, P_T) + \lambda^*,
\end{equation}

\noindent where $\epsilon_T$ and $\epsilon_S$ denote the expected prediction error on the target and source domains respectively, $\mathcal{D}_{\mathcal{H}}$ measures the divergence between the two distributions in hypothesis space $\mathcal{H}$, and $\lambda^*$ is the irreducible error of the optimal joint hypothesis. By updating $h_u$ locally to minimize $\epsilon_T$, the PAP mechanism reduces the distribution gap $\mathcal{D}_{\mathcal{H}}$ and improves the personalization capacity of the model.

\subsection{Theoretical Analysis}

Next, we present theoretical analysis about model effectiveness and privacy guarantee by federated adapter tuning.

\subsubsection{Model Effectiveness.} In FRs, all clients collaboratively learn a shared item representation model $\varphi$ by minimizing:
\[
\min_{\varphi} \frac{1}{n} \sum_{u=1}^n \ell_u(\varphi)
\]
where $\ell_u$ is the local loss for the $u$-th client. However, this global optimization may lead to sub-optimal performance in heterogeneous settings due to client distribution shifts.

We adopt a personalized formulation where clients jointly learn a shared item representation model $\varphi$, while locally adapting their own adapters $\psi_u$, the revised objective is:
\[
\min_{\psi_u, \varphi} \frac{1}{n} \sum_{u=1}^n \min_{\psi_u \in \mathcal{H}} \ell_u(\psi_u \circ \varphi)
\]
This scheme allows better personalization and representation alignment for downstream recommendation. Under a linearized setting, we provide the following result:

\begin{theorem}[Effectiveness of Personalized Adapters]\label{thm:effe}
Let the global item representation model be a linear transformation $\mathbf{A} \in \mathbb{R}^{L \times f}$ and the local adapters for client $u$ be $\mathbf{B}_u \in \mathbb{R}^{f \times d}$. Then, we have:
\begin{itemize}
    \item The optimal solution to the personalized formulation
    \[
    \min_{\mathbf{A}, \{\mathbf{B}_u\}} \frac{1}{n} \sum_{u=1}^n \| \mathbf{A} \mathbf{B}_u - \mathbf{A}^* \mathbf{B}_u^* \|_2^2
    \]
    achieves zero loss when $\mathbf{A}$ spans the same space as $\mathbf{A}^*$ and $\{\mathbf{B}_u^*\}$ spans $\mathbb{R}^d$.

    \item The standard formulation with a shared global model
    \[
    \min_{\mathbf{A}, \mathbf{B}} \frac{1}{n} \sum_{u=1}^n \| \mathbf{A} \mathbf{B} - \mathbf{A}^* \mathbf{B}_u^* \|_2^2
    \]
    incurs residual error increasing with the heterogeneity of $\{\mathbf{B}_u^*\}$.
\end{itemize} 
\end{theorem}

\subsubsection{Privacy Guarantee.} We analyze the privacy benefits of our proposed federated adapter tuning in FR settings through an information-theoretic lens. Our key findings are summarized as follows.

\begin{theorem}[Mutual Information Upper Bound]
\label{thm:mi-bound}
Let $\psi_u$ be the adapter parameters trained on user data $\mathcal{D}_u$, with bounded value range $[-C, C]$ and quantization resolution $\delta$. The information leakage $I(\psi_u; \mathcal{D}_u)$ is upper bounded by:
\[
I(\psi_u; \mathcal{D}_u) \le fd \cdot \log\left( \frac{2C}{\delta} \right)
\]
where $d$ is the bottleneck dimension in the adapter.
\end{theorem}

\begin{theorem}[Mutual Information Bound from DP Mechanism]
\label{thm:dp-bound}
If Gaussian noise is added to adapter parameters $\tilde{\psi_u}$ such that the update satisfies $(\epsilon, \delta)$-differential privacy (DP), then the mutual information leakage is bounded by:
\[
I(\tilde{\psi_u}; \mathcal{D}_u) \le \epsilon^2 + \delta
\]
\end{theorem}

These results suggest that federated adapter tuning leaks significantly less information compared to full-model or embedding updates, due to its structural bottleneck and optional DP perturbation.

\section{Experiments}

In this section, we conduct experiments to answer the following research questions:

\textbf{RQ1:} Does our proposed TransFR method outperform the state-of-the-art methods regarding transferability?

\textbf{RQ2:} Does our proposed TransFR surpass existing transferable baselines in cold-start settings?

\textbf{RQ3:} How do the specific components in our proposed method work for FR tasks, \textit{i.e.,} pre-training, federated adapter-tuning, and post-adaptation personalization?

\textbf{RQ4:} How do certain hyperparameters within our method impact the FR tasks, \textit{i.e.,} the number of layers in the adapter, the number of epochs in post-adaptation personalization?

\subsection{Experimental Settings}

\subsubsection{Datasets \& Metrics}

To benchmark our model against baselines, we perform experiments with the \textit{Amazon}~\cite{ni2019justifying} dataset. We selected three  categories as distinct domains: Movies, Music, and Books. Within each domain, we excluded interaction records lacking titles of items and applied filters to retain users with at least 20 interactions and items with a minimum of 30 interactions, in alignment with previous studies~\cite{ncf_2017,zhao2020catn}. To assess our model and baselines, we employ a leave-one-out evaluation approach and two prevalent evaluation metrics in ranking tasks: HR and NDCG. 




\subsubsection{Comparison Models}

We adopt two kinds of benchmarks for overall comparisons, \textit{i.e.,} centralized models, i.e., CMF~\cite{singh2008relational}, EMCDR ~\cite{embeddingcr_2021}, DisenCDR~\cite{disencdr2022}, and recent federated approaches, i.e., FCF~\cite{fcf_2019}, IFedRec~\cite{ifedrec2024}, FELLRec~\cite{fellrec}, FedCT~\cite{fedct_2021} and FedDCSR~\cite{zhang2023feddcsr}. 





\subsection{Experimental Results}

In this section, we detail the experimental results on Amazon dataset to verify the effectiveness of the proposed model.


\subsubsection{Overall Performance}

We compare centralized and federated models, with the first three methods being centralized and the last five federated. Most rely on ID-based item embeddings and perform knowledge transfer via explicit ID mapping. To address RQ1, we evaluate the transferability of our method and baselines, as shown in Table~\ref{tab:transhr} and Table~\ref{tab:transndcg}, reporting HR@10 and NDCG@10 on Amazon dataset. Transferability is measured by $\Delta = \frac{\text{Source} - \text{Target}}{\text{Source}}$, using HR@10 as an example. The results reveal several key insights to generalize across domains.

\begin{table*}[t]
\renewcommand\arraystretch{1.2}
\footnotesize
\begin{center}
  \setlength\tabcolsep{3pt}
  \begin{tabular}{lccc|ccc|ccc|ccc}
    \toprule
     \multirow{2}{*}{\textbf{Models}} &  \multicolumn{3}{c}{\textbf{Music}$\to$\textbf{Books}}  & \multicolumn{3}{c}{\textbf{Music}$\to$\textbf{Movies}}  & \multicolumn{3}{c}{\textbf{Books}$\to$\textbf{Movies}}  &\multicolumn{3}{c}{\textbf{Average}} \\ \cline{2-13}
          & \textbf{Source} $\uparrow$ & \textbf{Target} $\uparrow$ & $\Delta$ $\downarrow$ & \textbf{Source} $\uparrow$ & \textbf{Target} $\uparrow$ & $\Delta \downarrow$ &\textbf{Source}$\uparrow$& \textbf{Target}$\uparrow$& $\Delta\downarrow$ &\textbf{Source}$\uparrow$& \textbf{Target}$\uparrow$& $\Delta\downarrow$  \\
    \midrule
\textbf{CMF}  &  0.2232&	0.1833&	17.88\%	&0.2245	&0.1378	&38.62\%	&0.1792& 	0.1455	&18.81\%&	0.2090& 	0.1555& 	25.57\% \\
\textbf{EMCDR} &  0.2232	&0.2149&	3.72\%&	0.2245&	0.1893&	15.68\%&	0.1792 &	0.1652	&7.81\%	&0.2090 &	0.1898 &	9.17\%
  \\
\textbf{DisenCDR} & 0.2404  & 0.2243  & 6.70\%  & 0.2302 & 0.1988 & 13.61\% & 0.2105 & 0.1912 & 9.17\% & 0.2270 & 0.2048 & 9.49\% \\
\hline
\textbf{FCF}       & 0.1533 &	0.1380 &	9.98\% &	0.1803	 &0.1154 &	36.00\%	 &0.1220  &	0.1004 &	17.70\%	 &0.1519  &	0.1179  &	22.34\% \\
\textbf{IFedRec} & 0.2193 & 0.1981 & 9.66\% & 0.2095 & 0.1810 & 13.61\% & 0.1912 & 0.1703 & 10.95\% & 0.2067 & 0.1831 & 11.40\% \\
\textbf{FELLRec} & 0.2215 & 0.2003 & 9.56\% & 0.2121 & 0.1844 & 13.06\% & 0.1939 & 0.1731 & 10.76\% & 0.2092 & 0.1859 & 11.13\% \\
\textbf{FedCT} & 0.2240 & 0.2034 & 9.20\% & 0.2142 & 0.1765 & 17.60\% & 0.1963 & 0.1585 & 19.24\% & 0.2115 & 0.1795 & 15.35\% \\
\textbf{FedDCSR}    &  \textbf{0.2417} & 	0.2217 & 	8.27\% & 	0.2221 & 	0.1881 & 	15.31\%	 & 0.1994  & 	0.1719	 & 13.79\%	 & 0.2211  & 	0.1939  & 	12.29\%\\
\hline
\textbf{TransFR}  & 0.2386  & \textbf{0.2335}  &  \textbf{2.14\%} & \textbf{0.2311}  & \textbf{0.2123}  & \textbf{8.14\%}   &\textbf{0.2146}  &\textbf{0.2089} & \textbf{2.66\%}  & \textbf{0.2281} & \textbf{0.2182} & \textbf{4.33\%} \\
    \bottomrule
  \end{tabular}
  \end{center}
    \caption{The results of our TransFR and the baselines in terms of transferability towards HR@10 on Amazon dataset. $\Delta$ represents the transferability capability from the source domain to the target domain. The best results in federated settings are in bold. $\uparrow$ indicates that higher values correspond to better performance while $\downarrow$ denotes the lower values, the better performance.}
  \label{tab:transhr}
\end{table*}

\begin{table*}[t]
\renewcommand\arraystretch{1.2}
\footnotesize
\begin{center}
  \setlength\tabcolsep{3pt}
  \begin{tabular}{lccc|ccc|ccc|ccc}
    \toprule
     \multirow{2}{*}{\textbf{Models}} &  \multicolumn{3}{c}{\textbf{Music}$\to$\textbf{Books}}  & \multicolumn{3}{c}{\textbf{Music}$\to$\textbf{Movies}}  & \multicolumn{3}{c}{\textbf{Books}$\to$\textbf{Movies}}  &\multicolumn{3}{c}{\textbf{Average}} \\ \cline{2-13}
          & \textbf{Source} $\uparrow$ & \textbf{Target} $\uparrow$ & $\Delta$ $\downarrow$ & \textbf{Source} $\uparrow$ & \textbf{Target} $\uparrow$ & $\Delta \downarrow$ &\textbf{Source}$\uparrow$& \textbf{Target}$\uparrow$& $\Delta\downarrow$ &\textbf{Source}$\uparrow$& \textbf{Target}$\uparrow$& $\Delta\downarrow$  \\
    \midrule
     \textbf{CMF}  &  0.1392	 & 0.0955 & 	31.39\%	 & 0.1132 & 	0.0764 & 	32.51\%	 & 0.0921  & 	0.0662 & 	28.12\%	 & 0.1148 	 & 0.0794 	 & 30.89\%
 \\
     \textbf{EMCDR} &  0.1392 & 	0.1121 & 	19.47\% & 	0.1132	 & 0.0901	 & 20.41\% & 	0.0921 &  	0.0705	 & 23.45\%	 & 0.1148  & 	0.0909  & 	20.84\%
  \\
\textbf{DisenCDR} & 0.1403 & 0.1182 & 15.75\% & 0.1121 & 0.0923 & 17.66\% & 0.0995 & 0.0821 & 17.48\% & 0.1173 & 0.0975 & 16.87\% \\
    \hline
    \textbf{FCF}       & 0.0982	  & 0.0663  & 	32.48\%	  & 0.0721  & 	0.0583  & 	19.14\%  & 	0.0798   & 	0.0421	  & 47.24\%	  & 0.0834   & 	0.0556   & 	33.35\%
 \\
\textbf{IFedRec} & 0.1010 & 0.0834 & 17.43\% & 0.0954 & 0.0673 & 29.45\% & 0.0831 & 0.0497 & 40.19\% & 0.0932 & 0.0668 & 28.30\% \\
\textbf{FELLRec} & 0.1145& 0.0995& 13.10\% &	0.1068& 0.0812& 23.97\% &	0.1033& 0.0831& 19.55\%&	0.1082& 0.0879& 18.76\% \\
\textbf{FedCT} & \textbf{0.1221}&  0.1122&  8.11\%& 	\textbf{0.1144}&  0.0899&  21.41\%&	0.1047&  0.0845&  19.29\%	& \textbf{0.1137}&  0.0955&  16.01\% \\
    \textbf{FedDCSR}    & 0.1192 & 	0.1085 & 	8.98\%	 & 0.1143	 & 0.0892	 & 21.96\%	 & 0.0943  & 	0.0732 & 	22.38\% & 	0.1093 &  	0.0903  & 	17.36\%
 \\
    \hline
    \textbf{TransFR}  & 0.1209  & \textbf{0.1133}  &  \textbf{6.29\%} & 0.1132 & \textbf{0.0934}  & \textbf{17.49\%}   &\textbf{0.1054}  &\textbf{0.0893} & \textbf{15.28\%}  & 0.1132 & \textbf{0.0987} & \textbf{12.81\%} \\
    \bottomrule
  \end{tabular}
  \end{center}
    \caption{The results of our TransFR and the baselines in terms of transferability towards NDCG@10 on Amazon dataset.}
  \label{tab:transndcg}
\end{table*}


Centralized approaches can achieve better transferability, but they inevitably pose privacy risks. Among federated baselines, FedDCSR and FedCT outperforms FCF, IFedRec and FELLREC. This advantage stems from its use of explicit mapping to extract domain-relevant features. In contrast, FCF, IFedRec and FELLRec directly apply source-trained models to the target domain. IFedRec performs better than FCF due to its textual side information, while FCF relies on matrix factorization with sparse interaction data.

Our TransFR method achieves the best target domain performance, with the lowest $\Delta$ values in HR@10 (4.33\%) and NDCG@10 (12.81\%), demonstrating superior transferability. Although it slightly underperforms FedCT and FedDCSR in some source domain cases, it consistently outperforms them in target domain metrics and transfer loss. This advantage arises from using universal textual representations that reduce semantic gaps across domains, unlike ID-based methods relying on explicit ID mappings. 


\begin{table}[t]
\renewcommand\arraystretch{1.2}
\small
\begin{center}
  \setlength\tabcolsep{0pt}
  \begin{tabular}{lcc|cc|cc}
    \toprule
     \multirow{2}{*}{\textbf{Models}} &  \multicolumn{2}{c}{\textbf{Music}$\to$\textbf{Books}}  & \multicolumn{2}{c}{\textbf{Music}$\to$\textbf{Movies}}  & \multicolumn{2}{c}{\textbf{Books}$\to$\textbf{Movies}}   \\ \cline{2-7}
          & \textbf{H@10} $\uparrow$ & \textbf{N@10} $\uparrow$ & \textbf{H@10} $\uparrow$ & \textbf{N@10} $\uparrow$ & \textbf{H@10} $\uparrow$ & \textbf{N@10} $\uparrow$ \\
    \midrule
     \textbf{CMF} &  0.1833  & 0.0955  & 0.1378 & 0.0764  & 0.1455 & 0.0662   \\
     \textbf{EMCDR}  &  0.2149  & 0.1121  & 0.1893 & 0.0901  & 0.1652 & 0.0705 \\
     \textbf{DisenCDR}  & 0.2287   & 0.1123  & 0.1934 & 0.0921  & 0.1892 & 0.0764 \\
    \hline
    \textbf{FCF}      & 0.1380  & 0.0663  & 0.1154  & 0.0583  & 0.1004  & 0.0421   \\
    \textbf{IFedRec}  & 0.1944   & 0.0982   & 0.1654 & 0.0631   & 0.1583 & 0.0681 \\
    \textbf{FELLRec}  & 0.2093   & 0.1052  & 0.1723 & 0.0845  & 0.1639 & 0.0701 \\
    \textbf{FedCT}  &  0.2138   & 0.1062 & 0.1802 & 0.0858  & 0.1708 & 0.0726 \\
    \textbf{FedDCSR}    &  0.2217  & 0.1085  & 0.1881  & 0.0892  & 0.1719 & 0.0732   \\
    \hline
    \textbf{TransFR}  & \textbf{0.2310}  & \textbf{0.1126}  &  \textbf{0.2001} & \textbf{0.0934}  & \textbf{0.1932}  & \textbf{0.0830}   \\
    \bottomrule
  \end{tabular}
  \end{center}
    \caption{Comparison results of TransFR and the baselines on Amazon dataset in cold-start settings.}
  \label{tab:coldstart}
\end{table}

Table~\ref{tab:coldstart} aims to address RQ2, reporting HR@10 and NDCG@10 on Amazon dataset in cold-start settings, where models are evaluated on unseen items in the target domain. We derive several key observations. First, the federated baseline FCF struggles due to its ID-based item representation. IFedRec offers slight improvement by adding textual data but still relies on IDs, limiting its effectiveness. FedDCSR uses disentangled representations to transfer knowledge across domains, partially alleviating cold-start issues. Our method surpasses both ID-based and ID-mapping baselines by leveraging domain-agnostic textual embeddings and federated adapter tuning. Unlike coarse-grained ID embeddings, our textual representations capture item semantics more effectively and remain consistent across domains.


\begin{figure}[t]
\begin{center}
\includegraphics[scale=0.34]{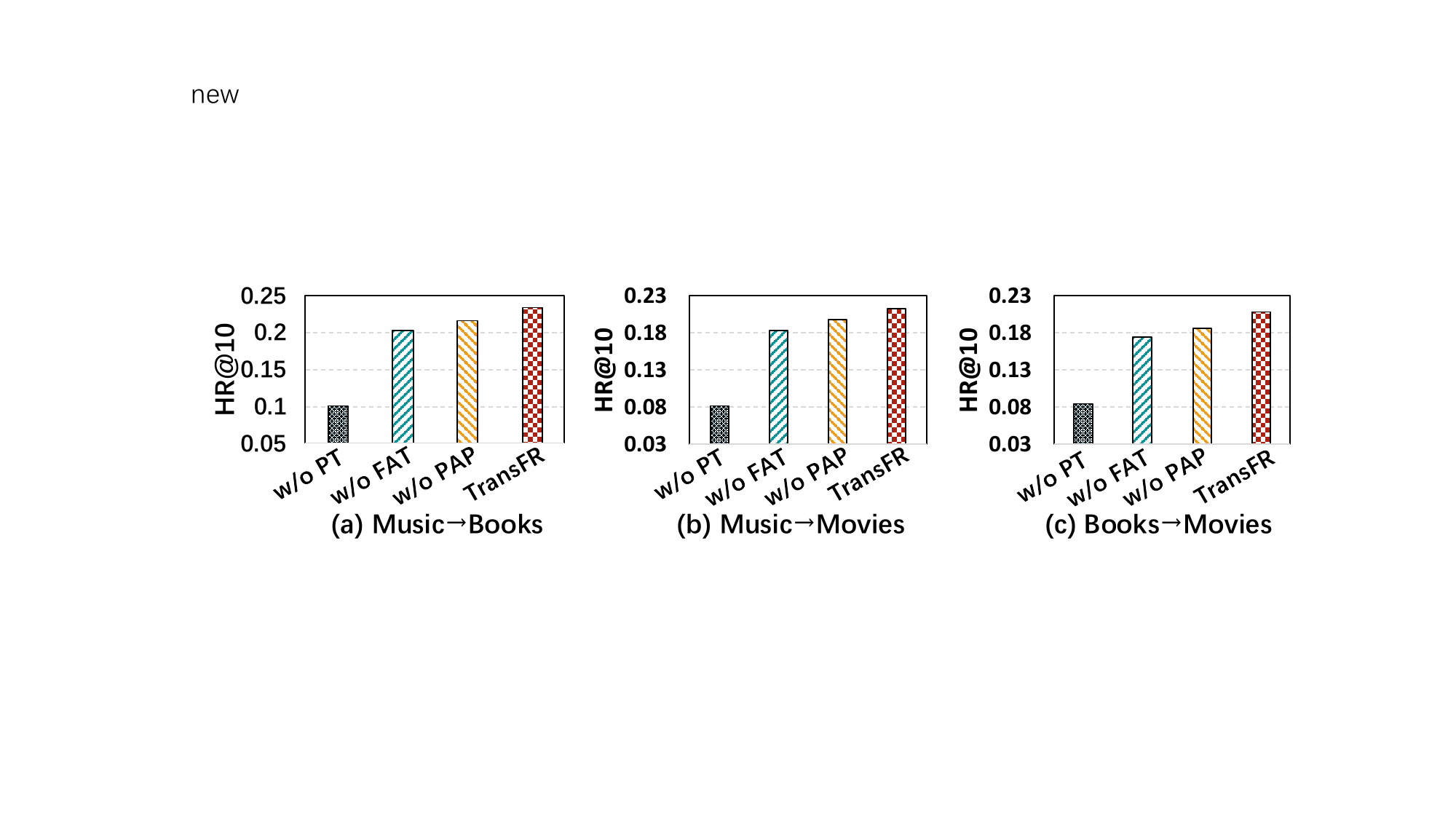}
\end{center}
\caption{Ablation study results towards HR@10 in the three transferable tasks on Amazon dataset.
}
\label{fig:hr_ablation}
\end{figure}

\subsubsection{Ablation Study}

\begin{figure}[h]
\begin{center}
\includegraphics[scale=0.32]{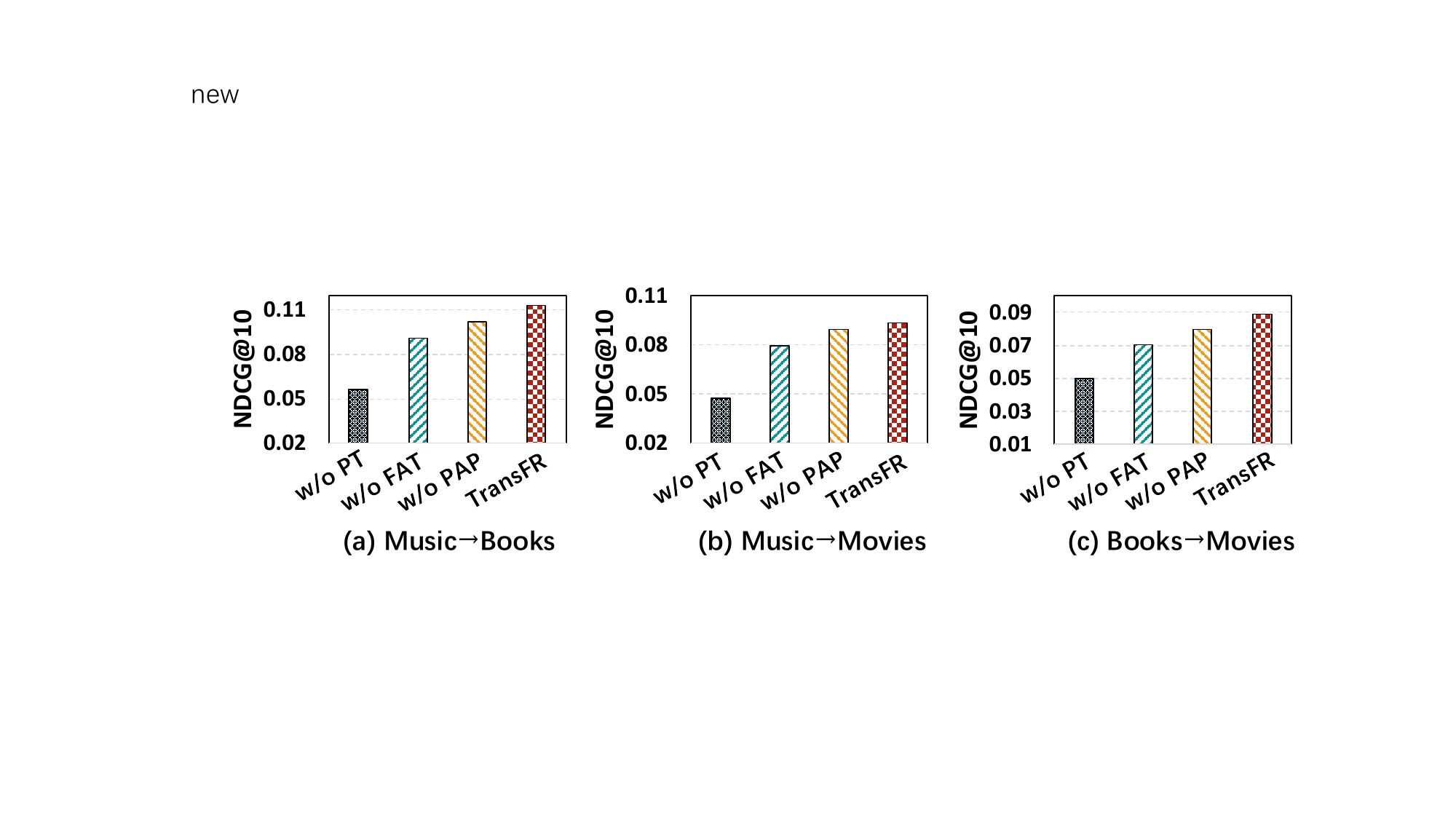}
\end{center}
\caption{Ablation study results towards NDCG@10 in the three transferable tasks on Amazon dataset.
}
\label{fig:ndcg_ablation}
\end{figure}

To address RQ3, we conduct an ablation study to assess the contribution of each component, i.e., pre-training (PT), federated adapter-tuning (FAT), and post-adaptation personalization (PAP), as shown in Fig.\ref{fig:hr_ablation} and Fig.\ref{fig:ndcg_ablation}. We compare three variants, each removing one module: “w/o PT” replaces pre-training with random initialization, “w/o FAT” omits federated adapter tuning, and “w/o PAP” removes local head adaptation. As shown in Fig.\ref{fig:hr_ablation} and Fig.\ref{fig:ndcg_ablation}, we observe the following: (1) Removing pre-training significantly degrades performance, aligning with prior findings~\cite{chen2022importance}; (2) Federated adapter-tuning yields moderate gains, validating the effectiveness of our Adapter module; (3) Post-adaptation personalization further boosts transferability, highlighting the role of the local head in personalized recommendation.

\begin{table}[h]
\renewcommand\arraystretch{1.2}
\small
\begin{center}
  \setlength\tabcolsep{1.8pt}
  \begin{tabular}{lcc|cc|cc}
    \toprule
     \multirow{2}{*}{ $\delta$} &  \multicolumn{2}{c}{\textbf{Music}$\to$\textbf{Books}}  & \multicolumn{2}{c}{\textbf{Music}$\to$\textbf{Movies}}  & \multicolumn{2}{c}{\textbf{Books}$\to$\textbf{Movies}}   \\ \cline{2-7}
         & \textbf{H@10} $\uparrow$ & \textbf{N@10} $\uparrow$ & \textbf{H@10} $\uparrow$ & \textbf{N@10} $\uparrow$ & \textbf{H@10} $\uparrow$ & \textbf{N@10} $\uparrow$ \\
    \midrule
     0.0 &  0.2310  & 0.1126  & 0.2001 & 0.0934  & 0.1932 & 0.0830  \\
     0.1   & 0.2301  & 0.1123 & 0.1989 & 0.0928  & 0.1913 & 0.0802  \\ 
     0.2 &  0.2289  & 0.1101  & 0.1964 & 0.0903  & 0.1889 & 0.0775   \\
     0.3  &  0.2234  & 0.1089  & 0.1903 & 0.0875  & 0.1845 & 0.0734  \\
    \bottomrule
  \end{tabular}
  \end{center}
    \caption{Results of TransFR with differential privacy.}
  \label{tab:dp}
\end{table}

We further incorporate the differential privacy to enhance privacy capability, with $\delta$ denoting noise intensity into adapters, which leads to only a slight performance decline, as shown in Table~\ref{tab:dp}. This shows that TransFR maintains competitive performance while keeping privacy protection.

\begin{table}[htbp]
\renewcommand\arraystretch{1.0}
\small
\begin{center}
  \setlength\tabcolsep{0.2pt}
  \begin{tabular}{lcc|c|c|c}
    \toprule
     \multirow{2}{*}{\textbf{Models}} &  \multicolumn{2}{c}{\textbf{Accuracy}}  & \multicolumn{1}{c}{\textbf{Memory}}  & \multicolumn{1}{c}{\textbf{Commu.}} & \multicolumn{1}{c}{\textbf{Inference}}   \\ \cline{2-6}
          & \textbf{H@10} $\uparrow$ & \textbf{N@10} $\uparrow$ & \textbf{MB} $\downarrow$ & \textbf{MB} $\downarrow$ & \textbf{s} $\downarrow$  \\
    \midrule
     \textbf{DistBERT}  &  0.2182  & 0.0987  & \textbf{68}   & \textbf{68} &  \textbf{0.003}  \\
     \textbf{Llama-2-7B}    & \textbf{0.2241}  & \textbf{0.1014} &  14,336  & 14,336 &  0.015 \\
     \hline
     \textit{\textbf{Gain}}    & 2.70\%  & 2.73\% &  $-$20,982\%  & $-$20,982\% &  $-$400\% \\ 
    \bottomrule
  \end{tabular}
  \end{center}
    \caption{Ablation study comparison of pre-trained modules between DistBERT and Llama-2 in terms of recommendation accuracy, memory overhead (MB), communication (commu.) overhead (MB), and inference time (s).}
  \label{tab:bert_llama}
\end{table}

\begin{figure}[htbp]
\begin{center}
\includegraphics[scale=0.33]{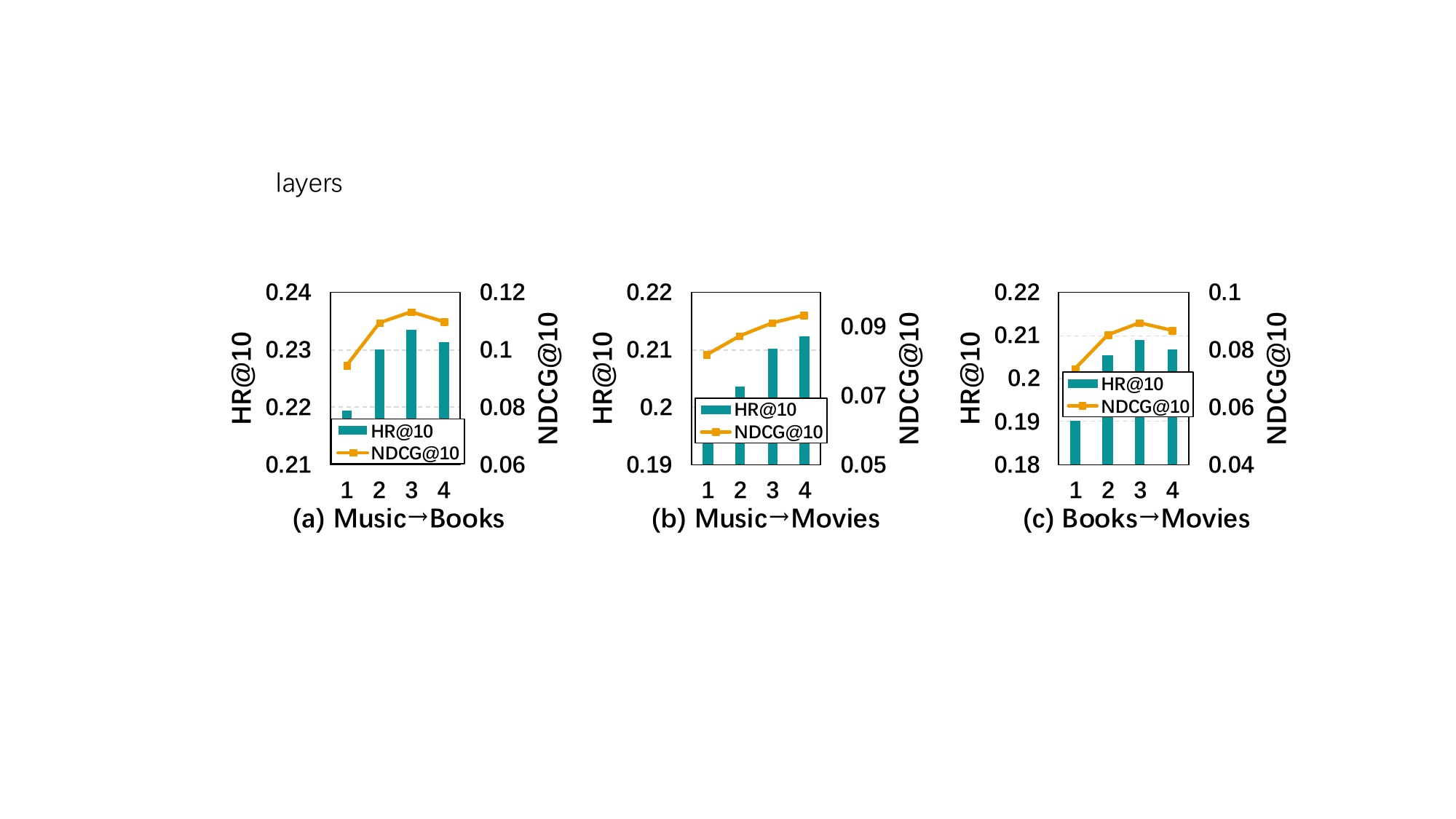}
\end{center}
\caption{Performance analysis of TransFR with varying numbers of layers in the adapter module.
}
\label{fig:local_layers}
\end{figure}

\begin{figure}[!htbp]
\begin{center}
\includegraphics[scale=0.33]{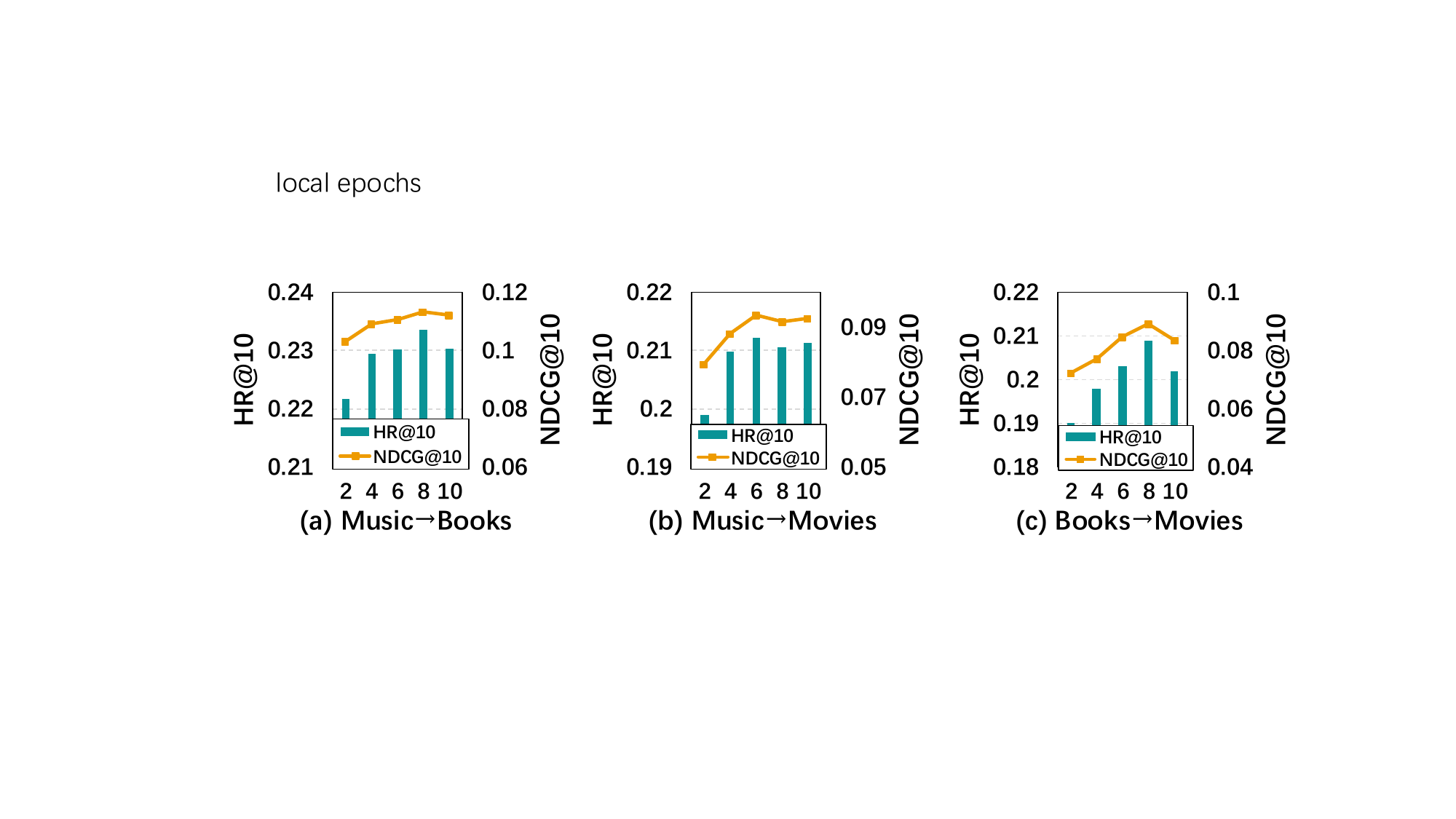}
\end{center}
\caption{Performance analysis of TransFR with different local epochs in post-adaptation personalization module.
}
\label{fig:local_traning}
\end{figure}

We evaluate the impact of different pre-trained models by comparing DistBERT and Llama-2-7B on Amazon dataset. As shown in Table~\ref{tab:bert_llama}, Llama-2-7B yields slightly better accuracy (+2.70\% H@10, +2.73\% N@10) than DistBERT, but at a high memory and communication overheads rise by 20,982\%, and inference cost increases by 400\%. These results suggest: (1) Large language models offer marginal accuracy gains but are inefficient for item embedding tasks. This finding is consistent with recent pioneer works~\cite{we1, we2}; (2) DistBERT provides a better trade-off between performance and efficiency for cross-device tasks.


\subsubsection{Sensitivity Analysis} To address RQ4, we analyze results with different hyperparameters. As shown in Fig.~\ref{fig:local_layers}, increasing the number of hidden layers generally improves performance in the target domain, though some tasks (e.g., Music → Books, Books → Movies) show diminishing returns after a certain point. Thus, a moderate increase in hidden layers can enhance cross-device recommendation performance. We further examine the effect of local training epochs in post-adaptation personalization on Amazon dataset, as shown in Fig.\ref{fig:local_traning}. Performance generally improves with more epochs, but overfitting may occur in scenarios with limited local data. Balancing effectiveness and efficiency, we set the number of local training epochs to 6 for all client devices.


\section{Conclusion}


In this work, we propose a transferable FR model with adapter tuning on pre-trained language models. By replacing domain-specific item embedding tables with domain-agnostic pre-trained textual representations and adapter tuning, we address the transferability, cold-start, and privacy issues. Besides, we discuss the superiority of TransFR on several aspects from theoretical and experimental perspectives. In future work, we plan to incorporate more powerful multi-modal foundation models to further enhance model performance within the FR framework.

\section{Acknowledgments}



This work was supported by the Beijing Natural Science Foundation 4264109, in part by the National Science Foundation of China under Grant U2268203, 62402031, 62203040, 62306027 and 62372436, and in part by the Beijing Nova Program 20240484620.

\bibliography{transfr}

\end{document}